\documentclass[11pt,a4paper]{article}

\usepackage{amsmath}
\usepackage{amsfonts}
\usepackage{amssymb}
\usepackage{epsfig}
\usepackage{graphics}

\usepackage{graphicx}

\begin{document}

\begin{titlepage}

\title{{\bf Stochastic Stabilization of Cosmological Photons}}

\vspace{2in}

\author{C.~P.\ Dettmann$^1$ J.~P.\ Keating$^1$ \& S.~D.\ Prado$^2$ \\ \\1.~School of Mathematics,\\ University of
Bristol,\\ Bristol BS8 1TW,\\ UK \\\\ and \\\\ 2.~Instituto de
F{\'\i}sica,\\ Universidade Federal do Rio Grande do Sul,\\ P.O.\
Box 15051,\\ 91501-970 Porto Alegre,\\ Brazil.}

\maketitle

\thispagestyle{empty}

\begin{abstract}
The stability of photon trajectories in models of the Universe
that have constant spatial curvature is determined by the sign of
the curvature: they are exponentially unstable if the curvature is
negative and stable if it is positive or zero. We demonstrate that random
fluctuations in the curvature provide an additional stabilizing
mechanism. This mechanism is analogous to the one responsible for
stabilizing the stochastic Kapitsa pendulum. When the mean
curvature is negative it is capable of stabilizing the photon
trajectories; when the mean curvature is zero or positive it
determines the characteristic frequency with which neighbouring
trajectories oscillate about each other.  In constant negative
curvature models of the Universe that have compact topology,
exponential instability implies chaos (e.g.~mixing) in the photon
dynamics.  We discuss some consequences of stochastic
stabilization in this context.
\end{abstract}

\vspace{5cm}


\maketitle

\end{titlepage}

\maketitle

One of the fundamental questions concerning the dynamical
properties of photons in the Universe is whether their
trajectories are stable or unstable. This strongly influences both
the images of distant objects as well as fluctuations in the
cosmic microwave background (CMB). In cosmological models in which
the spatial geometry of the Universe has constant curvature $K$
the photon trajectories (geodesics) are stable if $K>0$, in which
case neighbouring trajectories oscillate about each other with a
characteristic frequency determined by $K$, and exponentially
unstable if $K<0$, in which case neighbouring trajectories diverge
exponentially quickly with a Liapunov exponent determined by
$|K|$.

It is obvious, however, that the Universe is not exactly
homogeneous and isotropic: matter is not uniformly distributed but
is organized into galaxies, clusters of galaxies, and even
superclusters of galaxies \cite{peebles80}. Consequently, the
spatial curvature cannot be constant, but fluctuates.  Our purpose
here is to investigate the influence of these fluctuations on the
stability of the photon trajectories. It might be thought that
random fluctuations in the curvature would be a source of
instability. We show that this is not in fact the case; their
influence is a stabilizing one in that the curvature is
renormalized by an additional (positive) factor related to the
fluctuation amplitude and length-scales.  The mechanism
responsible for this {\it stochastic stabilization} is analogous
to the one that stabilizes the Kapitsa pendulum: a pendulum whose
pivot is forced to move periodically or stochastically along the
vertical \cite{kapitsa51}.

Given the fact that for constant curvature models to be consistent
with recent observations of the CMB the absolute value of the mean
curvature must be small -- the density of energy in the Universe
is within about $1\%$ of its critical value \cite{wmap03} --
stochastic stabilization is capable of dominating the stability
balance.  For example, an order of magnitude estimate suggests
that it is relevant in the present cosmological epoch.  The
difference between a constant curvature model with $K$ close to
zero and one in which stochastic stabilization dominates is
quantifiable in terms of the stability frequency with which
neighbouring trajectories oscillate about each other.

Our results apply to both open and compact geometries.  There has,
however, recently been considerable attention focused on negative
curvature models that have compact topology \cite{gurzadyan93,
lachieze95, cornish96, levin00, barrow01, aurich01, levin02}.
Photon dynamics in compact spaces is recurrent and hence, by
virtue of the exponential instability, strongly chaotic
(e.g.~mixing) when the curvature is constant and negative.
The fact that fluctuations in the cosmic microwave background
(CMB) have a distribution close to gaussian has been related to
Berry's random wave model \cite{berry77} for wave modes in chaotic
systems \cite{levin00, levin02} and the phenomenon of scarring in
these wave modes \cite{heller84} has been linked with anisotropic
structures in the CMB and in the distribution of galaxies
\cite{levin00, levin02}. We discuss some implications of
stochastic stabilization in this context.

We begin by deriving the appropriate form of the geodesic
deviation equation -- the equation that governs the separation of
neighbouring trajectories. Since the aim of this Letter is to
establish the principle of stochastic stabilization for photon
trajectories, the calculations we report contain the essential
ingredients for the effect, but ignore many additional,
comparatively weaker phenomena present in the early Universe. With
this in mind, we begin with an expanding isotropic homogeneous
(Friedmann) cosmology perturbed by density fluctuations with
non-relativistic velocities, neglecting perturbations of vector
and tensor character such as gravitational waves, and pressure
fluctuations, for example caused by relativistic neutrinos.  In
the coordinate system called the `conformal Newtonian' or
`longitudinal' gauge \cite{mukhanov92}, with the spatial variables
in Robertson-Walker form, such a spacetime is described by the
metric
\begin{multline}
\label{eq:metric} ds^2=R^2(\tau)\{(1+2\Phi)d\tau^2-
\frac{1-2\Phi}{[1+\frac{K}{4}(x^2+y^2+z^2)]^2}(dx^2+dy^2+dz^2)\}.
\end{multline}
Here $R(\tau)$ is the scale factor or spatial curvature radius of
the Universe, $\tau=\int R^{-1}dt$  is conformal time, $\Phi<<1$
is the Newtonian gravitational potential, $K$  is the
dimensionless spatial curvature (e.g.~$K=-1$ corresponds to a
hyperbolic geometry), $(x,y,z)$ are comoving coordinates expanding
at the same rate as the Universe, and units are chosen in which
Newton's constant $G$ and the speed of light $c$ are equal to
unity. The paths of photons through the spacetime are null
geodesics described by the equation~\cite{Seljak}
\begin{equation}
\frac{d{\bf n}}{d\tau}=-2\nabla_\perp\Phi
\end{equation}
where $\bf n$ is the photon direction, and $\nabla_\perp$ is the gradient
in comoving coordinates perpendicular to $\bf n$.  Note that since $\Phi$
and its derivative are small, the total change of $\bf n$ is small, and so
the transverse derivative can be replaced by the derivative transverse to
the observed (final) direction of the photon.  The separation of two closely
spaced photons propagates according to the geodesic deviation
equation~\cite{misner73} which under our approximations leads to
\begin{equation}
\label{eq:gde} \frac{d^2}{d\tau^2}\begin{pmatrix}
  \xi \\
  \eta
\end{pmatrix}=-\begin{pmatrix}
  K+2\Phi_{\xi\xi} & 2\Phi_{\xi\eta} \\
  2\Phi_{\xi\eta} & K+2\Phi_{\eta\eta}
\end{pmatrix}\begin{pmatrix}
  \xi \\
  \eta
\end{pmatrix}
\end{equation}
where $\xi$ and $\eta$ are the components of the separations orthogonal
to $\bf n$, and there are contributions arising from both the spatial
curvature $K$ and second derivatives of the gravitational potential
in directions orthogonal to $\bf n$, namely tidal forces.

These tidal forces are, in principle, entirely characterized by a
complete knowledge of the fluctuations in the matter density. We
shall consider them to be random functions of position (and
possibly time). As it moves under their influence, a photon will
be seen to be acted on by a time-varying force which fluctuates
rapidly (because the speed of light is large and the length scale
of the density fluctuations is small compared to $R$) and randomly
(i.e.~stochastically) with zero mean. We thus replace the
spatially random tidal force terms in (\ref{eq:gde}) with a
time-dependent stochastic perturbation.

In order to illustrate the qualitative behaviour of the solutions
of the resulting class of equations, we consider first the
analogous one-component case:
\begin{equation}
\label{eq:pend} \frac{d^2u}{d\tau^2}=-(k+Af(\tau))u.
\end{equation}
Here $A$ is a control parameter and $f(\tau)$ is a stochastic
forcing function, which we take to have zero mean. This has a
mechanical analogy: it describes a pendulum with a vertically
moving pivot in the limit of small oscillations. In this case $u$
denotes the angular displacement from the vertical and the forcing
term describes the height of the pivot. This problem was studied
in detail by Kapitsa \cite{kapitsa51}.  The constant gravitational
force represented by $k$ ($k<0$ corresponds to an up-turned
pendulum and $k>0$ to a down-turned one) plays the role of the
smooth geometry in the case of the unperturbed cosmology, while
the motion of the pivot corresponds to the metric perturbations
induced by fluctuations in the matter density.

We shall be interested in the case when the frequency $\omega$
characterizing the fluctuations in $f(\tau)$ is large.
Stabilization has been proved in the limit $\omega
\rightarrow\infty$ using Liapunov exponent techniques (see, for
example, \cite{kao94}).  The dependence of the stability on
$\omega$ can be deduced by separating $u$ asymptotically into fast
and slow components \cite{kapitsa51}: $u(\tau)=\langle
u(\tau)\rangle +u_f(\tau)$, where $\langle\ldots\rangle$ denotes a
local time average over scales large compared to $\omega^{-1}$.
Here $\langle u(\tau)\rangle$ is the slow component and, as
$\omega\rightarrow\infty$, $u_f$ is small and varies rapidly.
Substituting into (\ref{eq:pend}) and averaging, we have
\begin{equation}
\label{eq:pend1} \frac{d^2\langle u\rangle}{d\tau^2}=-k\langle
u\rangle -A\langle f(\tau)u_f(\tau)\rangle .
\end{equation}
Subtracting (\ref{eq:pend1}) from (\ref{eq:pend}) gives, as
$\omega \rightarrow\infty$,
\begin{equation}
\label{eq:pend2} \frac{d^2u_f}{d\tau^2}\approx -Af(\tau)\langle
u(\tau)\rangle .
\end{equation}
This equation can be integrated directly, treating $\langle
u(\tau)\rangle$ as a constant, leading to
\begin{equation}
\label{eq:pend3} \frac{du_f}{d\tau}\approx -A\langle
u(\tau)\rangle\int_{\tau_1}^{\tau}f(t)dt\equiv -A\langle
u(\tau)\rangle v(\tau),
\end{equation}
where $\tau_1$ is chosen so that $\langle v\rangle=0$, and
\begin{equation}
\label{eq:pend4} u_f(\tau)\approx -A\langle
u(\tau)\rangle\int_{\tau_2}^{\tau}v(t)dt\equiv -A\langle
u(\tau)\rangle x(\tau),
\end{equation}
$\tau_2$ being chosen so that $\langle x\rangle=0$.  Substituting
into (\ref{eq:pend1}) and computing the integral implicit in the
average by parts, we obtain
\begin{equation}
\label{eq:pendapprox} \frac{d^2\langle u\rangle}{d\tau^2}\approx
-(k+A^2\langle v^2(\tau)\rangle)\langle u\rangle.
\end{equation}

\begin{figure}
\includegraphics[angle=0,width=8.0cm,height=6cm]{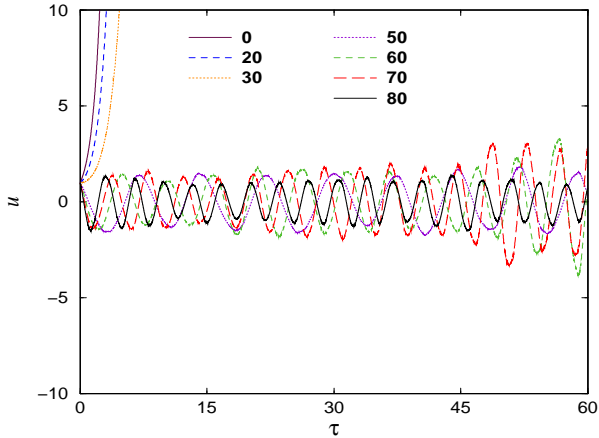}
\caption{Solutions of (\ref{eq:pend}) with $k=-1$ and $f(\tau)$
given by (\ref{eq:f}). When $A=0$, the solution increases
exponentially, corresponding to the fact that the photon
trajectories are unstable in a Universe with constant negative
curvature. This is also the case when $A=20$ and $A=30$.   When
$A=50, 60, 70$ and 80 the solution is oscillatory, indicating
stability.} \label{fig1}
\end{figure}

\begin{figure}
\includegraphics[angle=0,width=8.0cm,height=6cm]{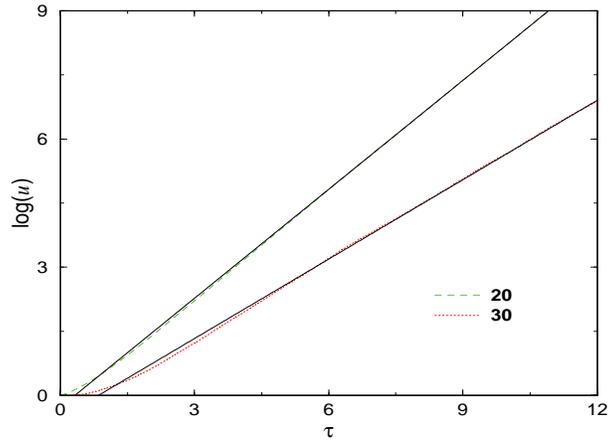}
\caption{Plots of $\log u$ vs.~$\tau$ for the data corresponding
to A=20 and A=30 in Figure \ref{fig1}.  The continuous black lines
are best-fitting straight lines when $\tau>6$.  For $A=20$ the
gradient is about 0.85 (the theoretical value is $0.8498\ldots$).
For $A=30$ the gradient is about 0.62 (the theoretical value is
$0.6124\ldots$).} \label{fig2}
\end{figure}

\begin{figure}
\includegraphics[angle=0,width=8.0cm,height=6cm]{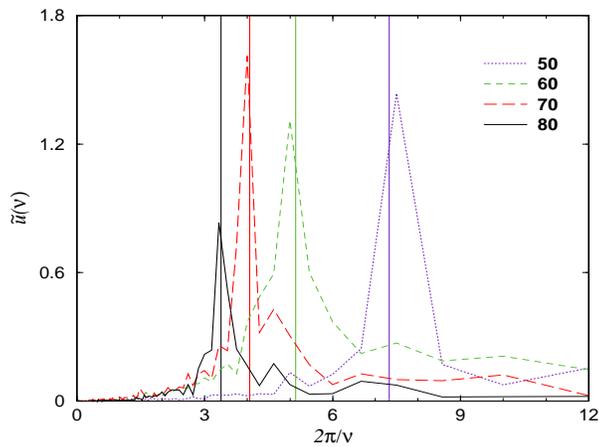}
\caption{Plots of the fourier transform of the data for $u(\tau)$
shown in Figure \ref{fig1}, as a function of period $2\pi/\nu$
(i.e.~$\nu$ is the frequency conjugate to $\tau$) when $A=50, 60,
70$ and 80. In each case the vertical dotted line marks the
theoretical oscillation period, $2\pi/\sqrt{A^2/1440-1}$.}
\label{fig3}
\end{figure}

It follows from (\ref{eq:pendapprox}) that if $A^2\langle
v^2\rangle<-k$ then $u$ grows exponentially as
$\tau\rightarrow\infty$, but if $A^2\langle v^2\rangle>-k$ then
$u$ is linearly stabilized.  This effect, and the accuracy of
(\ref{eq:pendapprox}), are illustrated by numerical simulations,
the results of which are represented in Figures \ref{fig1},
\ref{fig2} and \ref{fig3}.  In these simulations we took $k=-1$
and
\begin{equation}
\label{eq:f} f(\tau)=\sum_{n=1}^{N}\sin (\omega_n\tau +\phi_n)
\end{equation}
with $N=100$, frequencies $\omega_n$ chosen at random uniformly
from $[120, 600]$ and phases $\phi_n$ chosen at random uniformly
from $(-\pi,\pi ]$.  Thus $\langle v^2\rangle
=\frac{1}{2}\sum_n\frac{1}{w_n^2}\approx\frac{1}{1440}$ and so for
$A<\sqrt{1440}\approx 38$ the expectation is that $u$ grows like
$\exp (\tau\sqrt{1-A^2/1440})$, while for $A>\sqrt{1440}$ it
oscillates with frequency $\sqrt{A^2/1440-1}$.

The above analysis generalizes directly to the geodesic deviation
equation (\ref{eq:gde}).  This may be written as the stochastic
matrix equation
\begin{equation}
\label{eq:gdestoch} \frac{d^2}{d\tau^2}\begin{pmatrix}
  \xi \\
  \eta
\end{pmatrix}=-\begin{pmatrix}
  K+Af_{11}(\tau) & Af_{12}(\tau) \\
  Af_{12}(\tau) & K+Af_{22}(\tau)
\end{pmatrix}\begin{pmatrix}
  \xi \\
  \eta
\end{pmatrix}.
\end{equation}
The variables $\xi$ and $\eta$ can be separated into slow and fast
components, as for $u$. Following the steps that led from
(\ref{eq:pend}) to (\ref{eq:pendapprox}) we find, in the
high-frequency limit, that
\begin{equation}
\frac{d^2}{d\tau^2}\begin{pmatrix}
  \langle\xi\rangle \\
  \langle\eta\rangle
\end{pmatrix}\approx -(K\textbf{I}+A^2\langle \textbf{v}^2\rangle)\begin{pmatrix}
  \langle\xi\rangle \\
  \langle\eta\rangle
\end{pmatrix}
\end{equation}
where $\textbf{I}$ is the $2\times 2$ identity matrix and
$\textbf{v}$ is the matrix with elements
$v_{ij}=\int^{\tau}f_{ij}(t)dt$, such that $\langle
v_{ij}\rangle=0$. The stability of the trajectories thus depends
on the eigenvalues of the matrix
$K\textbf{I}+A^2\langle\textbf{v}^2\rangle$; specifically,
assuming $\langle v_{11}\rangle=\langle v_{22}\rangle$, they are
unstable if $A^2\langle(v_{11}\pm v_{12})^2\rangle<-K$, and if
$A^2\langle(v_{11}\pm v_{12})^2\rangle>-K$ they are stable.  Note
that $v_{11}$ and $v_{12}$ are of the order of $\omega^{-1}$, and
so the balance between the terms in the stability criterion
depends on the size of $A^2/\omega^2|K|$.  In particular, since
$\omega$ is considered large here, if $K<0$ the perturbations must
be large enough to cause the curvature to be positive in places
for stabilization to occur. This is consistent with the fact that
the geodesics on compact surfaces where the curvature is
everywhere negative (but not necessarily constant) are strongly
chaotic.

Let us make an order of magnitude estimate to determine the
cosmological epoch when the stochastic term is observable.  Since
the curvature is small, it will always be dominated by an
observable stochastic term (see below).  Consider that the
Universe consists of a volume fraction $(R_*/R)^3$ of clusters of
galaxies of size $L\approx 1$Mpc and mass $M\approx 10^{-4}$Mpc in
relativistic units.  The number of clusters within a ball of
radius $R$ is thus $\sim (R_*/L)^3$. The scale factor $R$ is set
to the scale of the visible Universe at present, roughly
$10^4$Mpc. $R_*$ is the distance scale at which voids between the
galaxies began to form, at a redshift of order 10 (i.e.~$R_*\sim
10^3$Mpc). The mean time (in terms of $\tau$) for a photon to move
between clusters is of the order of $LR^2/R_*^3$. This gives
$\omega\sim R_*^3/(LR^2)$, which is large up to the present, but
will fall to less than one in a few times the current age of the
Universe, as the galaxies move apart. Note that $\omega$ is
assumed large in the calculations described above.

The tidal forces $\Phi_{\xi\xi}$ are, up to factors of $4\pi$, of
the order of the dimensionless density $MR^2/L^3$, from Poisson's
equation when a photon passes near a cluster; $Av$ is typically
this quantity integrated over a mean free time, $MR^4/L^2R_*^3$.
$|K|\lesssim 0.01$ in these units.  The stochastic stabilization
will be observable if $Av\gtrsim 1$, and will dominate the
curvature if $Av/|K|\gtrsim 1$ (a weaker condition).  Taking this
together with the estimate for $\omega$, we conclude that
stochastic stabilization is potentially observable when
$(L^2/MR_*)^{1/4}\lesssim R/R_*\lesssim (R_*/L)^{1/2}$; that is,
for a period starting soon after the galaxies formed and due to
end a few times the present age of the Universe from now.
The magnitude of each deflection is determined by $M/L\sim 10^{-4}$, so
this effect should be observable on angular scales of arcseconds
or smaller~\cite{Seljak}.  From this perspective it should be easier
to observe in distant quasars, which are almost pointlike, than the
almost homogeneous cosmic background radiation.

The stochastic stabilization of photon trajectories has several
interesting consequences for cosmology.  In particular when
$K\approx 0$ it is observable in the characteristic frequency with
which neighbouring orbits oscillate about each other,
corresponding to a periodic focusing and defocusing of distant
images.

Given the recent focus on compact models of the Universe
\cite{levin02}, we devote some concluding remarks to this case
(although, as already noted, the mechanism is independent of
topology). The simplest models of the Universe that go beyond
having spatially constant negative curvature incorporate static
fluctuations. Such models are to a large extent still unrealistic,
because the curvature is unlikely to be static over the timescales
associated with photon recurrence. Nevertheless, they illustrate
most clearly the implications of stabilization, some of which we
now list.

As a result of stabilization, the photon dynamics will generically
possess both regular (stable) and irregular (chaotic) components,
rather than being fully chaotic. While wave modes in fully chaotic
systems are believed to have a gaussian value distribution, it is
well established in the context of quantum chaos that regular
trajectories lead to a quantifiable non-gaussian component whose
precise form depends on the size and position of the regular
regions in phase space \cite{backer02}.  This will in turn give
rise to a non-gaussian component in the CMB, via the connection
described in \cite{levin00, levin02}. There is, in addition,
likely to be a second non-gaussian component due to the photon
trajectories close to bifurcation (i.e.~close to making the
transition between instability and stability).  It was established
in \cite{keating01} that bifurcations give rise to non-gaussian
fluctuations in wave modes, quantified by characteristic scaling
exponents in the wavelength dependence of the moments of their
value distribution in the short-wavelength limit. In situations
when all of the generic bifurcations contribute, the moment
exponents take on universal values, which were calculated in
\cite{keating01}.  These non-gaussian statistics are also likely
to feed through to the fluctuations in the CMB in the models under
consideration here.

Bifurcations have a strong influence on the scarring of wave modes
as well.  In fully chaotic systems wave modes may be scarred by
periodic orbits \cite{heller84}.  This effect is dramatically
enhanced when the orbits undergo bifurcation \cite{keating01},
giving rise to superscars.  It has been suggested by Levin and
Barrow, in the context of constant negative curvature models, that
scars may give rise to anisotropic structures in the CMB
\cite{levin00, levin02}. This mechanism would therefore be
strongly enhanced by stabilization. Levin and Barrow
\cite{levin00, levin02} have also put forward the idea that the
anisotropic structures associated with scars may show up in the
distribution of matter. Regions of regular motion and superscars
would significantly amplify this effect.  In particular, stable
islands typically have a fractal distribution, and this may
provide an explanation, within the context of their theory, for
the fractal hierarchy of structures seen in galaxies and clusters
of galaxies. Stable orbits also strongly inhibit the rate of
mixing, giving rise to intermittency, and so may play a role in
the suggested links between mixing and preinflationary
homogenization \cite{cornish96}.









We gratefully acknowledge very helpful discussions with Dr Gopal
Basak and Professors Sir Michael Berry and Mark Birkinshaw.


\begin{thebibliography}{26}
\expandafter\ifx\csname
natexlab\endcsname\relax\def\natexlab#1{#1}\fi
\expandafter\ifx\csname bibnamefont\endcsname\relax
\def\bibnamefont#1{#1}\fi
\expandafter\ifx\csname bibfnamefont\endcsname\relax
\def\bibfnamefont#1{#1}\fi
\expandafter\ifx\csname citenamefont\endcsname\relax
 \def\citenamefont#1{#1}\fi
\expandafter\ifx\csname url\endcsname\relax
  \def\url#1{\texttt{#1}}\fi
\expandafter\ifx\csname
urlprefix\endcsname\relax\def\urlprefix{URL }\fi
\providecommand{\bibinfo}[2]{#2}

%

\providecommand{\eprint}[2][]{\url{#2}}



\bibitem
{peebles80}
 \bibinfo{author}{\bibfnamefont{P.~J.~E.}~\bibnamefont{Peebles}}, in
 \emph{\bibinfo{booktitle}{The Large-scale Structure of the Universe}}
    (\bibinfo{publisher}{Princeton Univ. Press, Princeton},
     \bibinfo{year}{1980}).



\bibitem
{kapitsa51}
\bibinfo{author}{\bibfnamefont{P.~L.}~\bibnamefont{Kapitsa}},
\bibinfo{journal}{Zh. Eksperim. Teor. Fiz.}
 \textbf{\bibinfo{volume}{21}}, \bibinfo{pages}{588} (\bibinfo{year}{1951}).



\bibitem
{wmap03}
\bibinfo{author}{\bibfnamefont{C.~L.} \bibnamefont{Bennett}} \bibnamefont{{\it et
al.}},
\bibinfo{journal}{Astrophys. J. Suppl. S.}
\textbf{\bibinfo{volume}{148}}, \bibinfo{pages}{1}
(\bibinfo{year}{2003}).



\bibitem
{gurzadyan93}
\bibinfo{author}{\bibfnamefont{V.~G.} \bibnamefont{Gurzadyan}} \bibnamefont{and}
  \bibinfo{author}{\bibfnamefont{A.~A.} \bibnamefont{Kocharyan}},
  \bibinfo{journal}{Astron. Astrophys.}
  \textbf{\bibinfo{volume}{260}}, \bibinfo{pages}{14} (\bibinfo{year}{1992}).



\bibitem
{lachieze95}
\bibinfo{author}{\bibfnamefont{M.}~\bibnamefont{Lachi\`eze-Rey}}
\bibnamefont{and}
\bibinfo{author}{\bibfnamefont{J.~P.}~\bibnamefont{Luminet}},
  \bibinfo{journal}{Phys. Rep.}
   \textbf{\bibinfo{volume}{254}}, \bibinfo{pages}{135} (\bibinfo{year}{1995}).



\bibitem
{cornish96}
\bibinfo{author}{\bibfnamefont{N.~J.} \bibnamefont{Cornish}},
 \bibinfo{author}{\bibfnamefont{D.~N.} \bibnamefont{Spergel}} \bibnamefont{and}
 \bibinfo{author}{\bibfnamefont{G.~D.} \bibnamefont{Starkman}}
 \bibinfo{journal}{Phys. Rev. Lett.}
 \textbf{\bibinfo{volume}{77}}, \bibinfo{pages}{215} (\bibinfo{year}{1996}).

\bibitem
{levin00}
  \bibinfo{author}{\bibfnamefont{J.}~\bibnamefont{Levin}} \bibnamefont{and}
  \bibinfo{author}{\bibfnamefont{J.~D.}~\bibnamefont{Barrow}},
     \bibinfo{journal}{Class. Quantum Grav.} \textbf{\bibinfo{volume}{17}},
     \bibinfo{pages}{L61} (\bibinfo{year}{2000}).



\bibitem
{barrow01}
\bibinfo{author}{\bibfnamefont{J.~D.}~\bibnamefont{Barrow}} \bibnamefont{and}
\bibinfo{author}{\bibfnamefont{J.}~\bibnamefont{Levin}},
  \bibinfo{journal}{Phys. Rev. A} \textbf{\bibinfo{volume}{63}},
          \bibinfo{pages}{044104} (\bibinfo{year}{2001}).



\bibitem
{aurich01}
\bibinfo{author}{\bibfnamefont{R.}~\bibnamefont{Aurich}} \bibnamefont{and}
\bibinfo{author}{\bibfnamefont{F.}~\bibnamefont{Steiner}},
  \bibinfo{journal}{Mon. Not. R. Astron. Soc.}
  \textbf{\bibinfo{volume}{323}}, \bibinfo{pages}{1016} (\bibinfo{year}{2001}).





\bibitem
{levin02}
\bibinfo{author}{\bibfnamefont{J.}~\bibnamefont{Levin}},
  \bibinfo{journal}{Phys. Rep.} \textbf{\bibinfo{volume}{365}},
  \bibinfo{pages}{251} (\bibinfo{year}{2002}).



\bibitem
{berry77}
\bibinfo{author}{\bibfnamefont{M.~V.} \bibnamefont{Berry}},
  \bibinfo{journal}{J.\ Phys. \ A} \textbf{\bibinfo{volume}{10}},
  \bibinfo{pages}{2083} (\bibinfo{year}{1977}).



\bibitem
{heller84}
\bibinfo{author}{\bibfnamefont{E.~J.}~\bibnamefont{Heller}},
\bibinfo{journal}{Phys. Rev. Lett.}
 \textbf{\bibinfo{volume}{53}}, \bibinfo{pages}{1515} (\bibinfo{year}{1984}).





\bibitem
{mukhanov92}
\bibinfo{author}{\bibfnamefont{V.~F.}~\bibnamefont{Mukhanov}},
 \bibinfo{author}{\bibfnamefont{H.~A.} \bibnamefont{Feldman}}
\bibnamefont{and} \bibinfo{author}{\bibfnamefont{R.~H.}~\bibnamefont{Brandenberger}},
\bibinfo{journal}{Phys. Rep.} \textbf{\bibinfo{volume}{215}},
 \bibinfo{pages}{203} (\bibinfo{year}{1992}).




\bibitem{Seljak}
\bibinfo{author}{\bibfnamefont{U.}~\bibnamefont{Seljak}},
	\bibinfo{journal}{Astrophys. J.}\textbf{\bibinfo{volume}{463}},
	\bibinfo{pages}{1} (\bibinfo{year}{1996}).

\bibitem
{misner73}
\bibinfo{author}{\bibfnamefont{C.~W.}~\bibnamefont{Misner}},
\bibinfo{author}{\bibfnamefont{K.~S.}~\bibnamefont{Thorne}} and
\bibinfo{author}{\bibfnamefont{J.~A.}~\bibnamefont{Wheeler}},
    \emph{\bibinfo{booktitle}{Gravitation
      }} (\bibinfo{publisher}{Freeman},
        \bibinfo{year}{1973}).



\bibitem
{kao94}
\bibinfo{author}{\bibfnamefont{J.} \bibnamefont{Kao}} \bibnamefont{and}
  \bibinfo{author}{\bibfnamefont{V.} \bibnamefont{Wihstutz}},
   \bibinfo{journal}{Stochastics and Stochastic Reports}
   \textbf{\bibinfo{volume}{49}}, \bibinfo{pages}{1} (\bibinfo{year}{1994}).







\bibitem
{backer02}
\bibinfo{author}{\bibfnamefont{A.} \bibnamefont{B\"acker}} \bibnamefont{and}
  \bibinfo{author}{\bibfnamefont{R.} \bibnamefont{Schubert}},
   \bibinfo{journal}{J. Phys.  A}
    \textbf{\bibinfo{volume}{35}}, \bibinfo{pages}{527} (\bibinfo{year}{2002}).



\bibitem
{keating01}
  \bibinfo{author}{\bibfnamefont{J.~P.} \bibnamefont{Keating}} \bibnamefont{and}
  \bibinfo{author}{\bibfnamefont{S.~D.} \bibnamefont{Prado}},
  \bibinfo{journal}{Proc.\ R.\ Soc.\ London, Ser.\ A}
  \textbf{\bibinfo{volume}{457}}, \bibinfo{pages}{1855} (\bibinfo{year}{2001}).









\end{thebibliography}
\end{document}